\documentclass[review]{elsarticle}

\usepackage{lineno,hyperref}
\journal{CARBON}

\begin{document}

\begin{frontmatter}

\title{Confined Electron and Hole States in Semiconducting Carbon Nanotube sub-10 nm Artificial Quantum Dots}

\author[csem]{Gilles Buchs\corref{cor1}}
\ead{gilles.buchs@csem.ch}

\author[DIPC,IKER]{Dario Bercioux\corref{cor2}}
\ead{dario.bercioux@dipc.org}

\author[FRAUNH1,FRAUNH2]{Leonhard Mayrhofer}

\author[EMPA]{Oliver Gr\"oning}

\cortext[cor1]{Corresponding author}
\cortext[cor2]{Corresponding author}

\address[csem]{CSEM, Swiss Center for Electronics and Microtechnology, Rue de l'Observatoire 58, CH-2000 Neuch\^{a}tel, Switzerland }
\address[DIPC]{Donostia International Physics Center (DIPC), Manuel de Lardizbal 4, 
E-20018 San Sebasti\'an, Spain}
\address[IKER]{IKERBASQUE, Basque Foundation for Science, Maria Diaz de Haro 3, 48013 Bilbao, Spain}
\address[FRAUNH1]{Fraunhofer IWM, W\"ohlerstra\ss e 11, D-79108 Freiburg, Germany}
\address[FRAUNH2]{Freiburg Materials Research Center FMF, University of Freiburg, Stefan-Meier-Strasse 21, D-79104 Freiburg, Germany}
\address[EMPA]{EMPA, Swiss Federal Laboratories for Materials Testing and Research, 
nanotech@surfaces, \"Uberlandstrasse 129, CH-8600 D\"ubendorf, Switzerland}

\begin{abstract}
We show that quantum confinement in the valence and conduction bands of semiconducting single-walled carbon nanotubes can be engineered by means of artificial defects. This ability holds potential for designing future nanotube-based quantum devices such as electrically driven, telecom-wavelength, room-temperature single-photon sources. Intrananotube quantum dots with sub-10 nm lateral sizes are generated between consecutive Ar$^{+}$ or N$^{+}$ ion-induced defects, giving rise to quantized electronic bound states with level spacings of the order of 100 meV and larger. Using low-temperature scanning tunneling spectroscopy, we resolve the energy and real space features of the quantized states and compare them with theoretical models. Effects on the states structure due to asymmetric defect scattering strength and the influence of the Au(111) substrate are remarkably well reproduced by solving the Schr\"odinger equation over a one-dimensional piecewise constant potential model. Using ab-initio calculations, we demonstrate that vacancies, chemisorbed nitrogen ad-atoms and highly stable double vacancies constitute strong scattering centers able to form quantum dots with clear signatures of discrete bound states as observed experimentally. The energy dependence of the defects scattering strength is also studied. Finally, steps toward a characterization of the optical properties of such quantum dot structures are discussed.
\end{abstract}


\end{frontmatter}

%\linenumbers

\section{Introduction}

Despite the rise of graphene and other 2D materials, semiconducting single-walled carbon nanotubes (SWNT) are still regarded as strong candidates for the next generation of high-performance ultrascaled transistors~\cite{Cao_IBM_2015,IBM_2017,3D_CNT_FET} as well as for opto-electronic devices~\cite{Review_Avouris,CNT_photonics} such as chip-scale electronic-photonic platforms~\cite{Pernice_2016} or low-threshold near-infrared tunable micro-lasers~\cite{Graf_2017}. 
Engineering a quantum dot (QD) along a (suspended) semiconducting SWNT foreshadows promising opportunities in the field of quantum information processing and sensing through recently proposed schemes such as detection and manipulation of single spins via coupling to vibrational motion~\cite{Palyi_2012}, optomechanical cooling~\cite{Wilson_Rae_2012} as well as all optical manipulation of electron spins~\cite{Galland_all_optical_2008}. Furthermore, the quasi one-dimensional geometry of SWNTs allows for defining tunable p-n junctions induced by electrostatic doping through local gates~\cite{Buchs_JAP,tunable_pn_2011}. Combining a well-defined QD within such a p-n junction structure could constitute a crucial building-block for the realization of highly desirable electrically driven, on-demand single photon emitters operating at telecom wavelength, based $e.g.$ on a turnstile device architecture~\cite{turnstile_1994,turnstile_1999}.
In practice, QDs in carbon nanotubes have been reported predominantly for two different confinement structures: i) Engineered tunneling barriers at metal-nanotube contacts~\cite{Pablo04nat} and/or by gate electrodes, used \emph{e.g.} to manipulate single electron spins~\cite{Laird:2015}, ii) Unintentional localization potentials stemming from environmental disorder~\cite{Hofmann_2016}, allowing for single-photon emission mediated by localization of band-edge excitons to QD states~\cite{CNT_photonics,Hoegele_2008,Walden_Newman_2012,Hofmann_2013,Pernice_2016_2}. Both types of structures are usually operated at cryogenic temperature due to small energy scales ranging from a few to few tens of millielectronvolts.
\\
\indent Another technique for achieving confinement in SWNTs makes use of artificial defects such as covalently bound oxygen or aryl functionalization groups on the side walls of semiconducting SWNTs, inducing deep exciton trap states allowing for single-photon emission at room temperature~\cite{Htoon_2015,tunable_QD_defects}. Also, carrier confinement between defect pairs acting as strong scattering centers has been reported for mechanically induced defects~\cite{Postma_SET} as well as for ion-induced defects with reported level spacings up to 200 meV in metallic SWNTs~\cite{Buchs_PRL}. The latter technique, combined with recent progress in controlling defects structure and localization~\cite{Robertson_2012,Yoon_2016,Laser_writing_2017} offers a high potential for engineering a broad set of SWNT-based quantum devices operating at room temperature. 
\\
\indent Here, we demonstrate confinement of electrons and holes in sub-10 nm QD structures defined by ion-induced defect pairs along the axis of semiconducting SWNTs. Using low temperature scanning tunneling microscopy and spectroscopy (STM/STS), bound states with level spacings of the order of 100 meV and larger are resolved in energy and space. By solving the one-dimensional Schr\"odinger equation over a piecewise constant potential model, the effects of asymmetric defect scattering strength as well as the influence of the Au(111) substrate such as terrace edges on the bound states structure are remarkably well reproduced. By means of ab-initio calculations based on density functional theory and Green's functions, we find that single (SV) and double vacancies (DV) as well as chemisorbed nitrogen ad-atoms are good candidates to produce QDs with the experimentally observed features. These simulations also allow to study the scattering profile as a function of energy for different defect combinations.

\section{Experimental section}

The experiments have been performed in a commercial (Omicron) low temperature STM setup operating at $\sim5$~K in ultra high vacuum. Topography images have been recorded in constant current mode with a grounded sample, using mechanically cut Pt/Ir tips. Differential conductance $dI/dV$ spectra, proportional in first approximation to the local density of states (LDOS)~\cite{Tersoff85} have been recorded using a lock-in amplifier technique. The LDOS spatial evolution along a nanotube axis is obtained by $dI/dV(x,V)$ maps built by a series of equidistant $dI/dV$ spectra. Spatial extent mismatches between topography images and consecutive $dI/dV(x,V)$ maps have been systematically corrected~\cite{Buchs_Ar}, and the metallic nature of the tip has been systematically checked on the gold substrate to prevent any tip artefacts before recording STM or/and STS data sets. 
\\
\indent Nanotube samples were made of extremely pure high-pressure CO conversion (HiPCo) SWNTs~\cite{Smalley01} with a diameter distribution centered around 1 nm, FWHM $\sim$ 0.3 nm~\cite{Buchs_conf}. The measured intrinsic defect density was below one defect every 200 nm. SWNTs were deposited on atomically flat Au(111) surfaces from a 1,2-dichloroethane suspension, followed by an in-situ annealing process~\cite{Buchs_APL_07,Buchs_Ar}.
\\
\indent Local defects in SWNTs have been created in-situ by exposure to: (i) Medium energy $\sim$ 200 eV argon ions (Ar$^{+}$) produced by an ion gun \cite{Buchs_Ar,Buchs_PRL}, (ii) Low energy (few eV's) nitrogen ions (N$^{+}$) produced by a 2.45 GHz ECR plasma source~\cite{Buchs_APL_07,Buchs_NJP_07}. In both cases, the exposure parameters have been calibrated to reach an average defect separation along the SWNTs of about 10 nm~\cite{Buchs_Ar,Buchs_APL_07}.

\section{Results and discussion}
\subsection{Experimental LDOS patterns}
%%%
%
\begin{figure}
  \includegraphics[width=8cm]{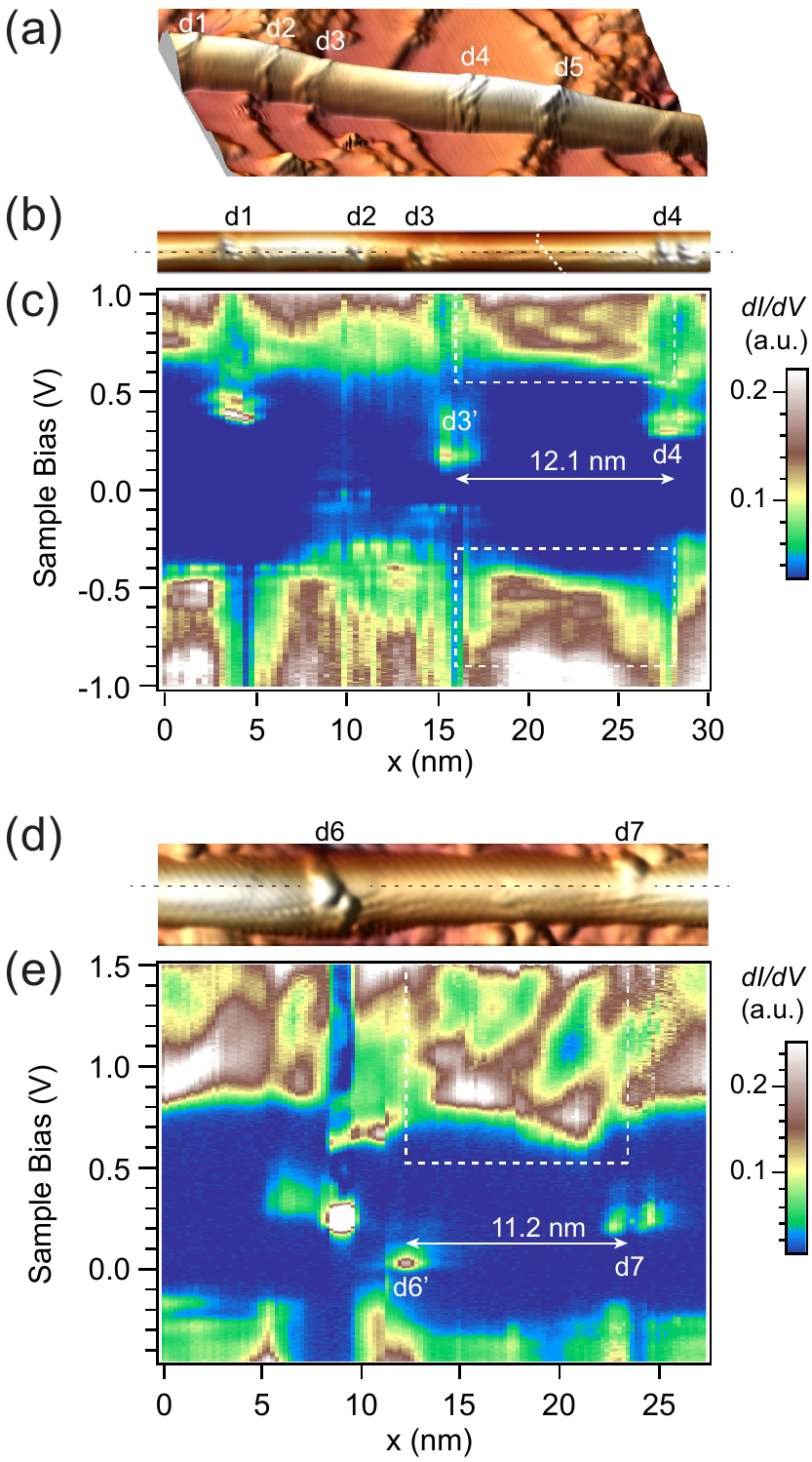}
  \caption{\label{exp_data_1} (a)-(b) 3D topography images (processed with WSXM~\cite{WSXM}) of SWNT I with Ar$^{+}$ ions-induced defects, with sample-tip bias voltage ($V_\mathrm{S}$) 1 V and tunneling current $I_\mathrm{S}$ 0.1 nA. (c) Corresponding $dI/dV(x,V)$ map recorded along the horizontal dashed lines in (b), with $V_\mathrm{S}=1$ V, $I_\mathrm{S}=0.2$ nA. Spatial resolution $\sim$ 0.3 nm. (d) 3D topography image of SWNT II with N$^{+}$ ions-induced defects, with $V_\mathrm{S}=1$ V, $I_\mathrm{S}=128$ pA. (e) Corresponding $dI/dV(x,V)$ map recorded along the horizontal dashed lines in (d), with $V_\mathrm{S}=1.5$ V, $I_\mathrm{S}=0.3$ nA. Spatial resolution $\sim$ 0.2 nm.}
\end{figure}
%
%%%
In Fig.~\ref{exp_data_1} (a) and (b), we show 3D STM images of the same semiconducting SWNT (referred as SWNT I in the following) with Ar$^{+}$ ions-induced defect sites labeled $d1-d5$ . Panel (d) shows a 3D STM image of a second semiconducting SWNT (referred as SWNT II) with N$^{+}$ ions-induced defect sites labeled $d6-d7$. In both cases, defect sites typically appear as hillock-like protrusions with an apparent height ranging from 0.5~{\AA} to 4~{\AA} and an apparent lateral extension varying between 5~{\AA} and 30~{\AA}~\cite{Buchs_NJP_07,Buchs_Ar,Thesis_Buchs}. 
\\
\indent The resulting $dI/dV(x,V)$ maps recorded along the horizontal dashed line drawn in the STM images (b) and (d) are displayed in panels (c) and (e) in Fig.~\ref{exp_data_1}, respectively. Defect signatures in the LDOS in both cases are characterized by deep in-gap states at the defects positions. This is consistent with the expected defect structures, $i.e.$ mainly SVs, DVs and combinations thereof for collisions with Ar$^{+}$ ions~\cite{Buchs_Ar} and bridgelike N ad-atom for collisions with N$^{+}$ ions~\cite{Thesis_Buchs,Nitrogen_prb_07}. Note that gap states at energy levels $\sim$~0.2 eV and $\sim$~0.05 eV in panels (c) and (e), respectively, are shifted to the right from $d3$ by about 1 nm and to the right from $d6$ by about 2 nm. This indicates the presence of intrinsic or ion-induced defects on the lateral or bottom side wall of the SWNTs~\cite{Kra01prb}, not visible in the topographic images. These defects are labelled $d3'$ and $d6'$, respectively.  
\\
%
%%%
\begin{figure}
  \includegraphics[width=12cm]{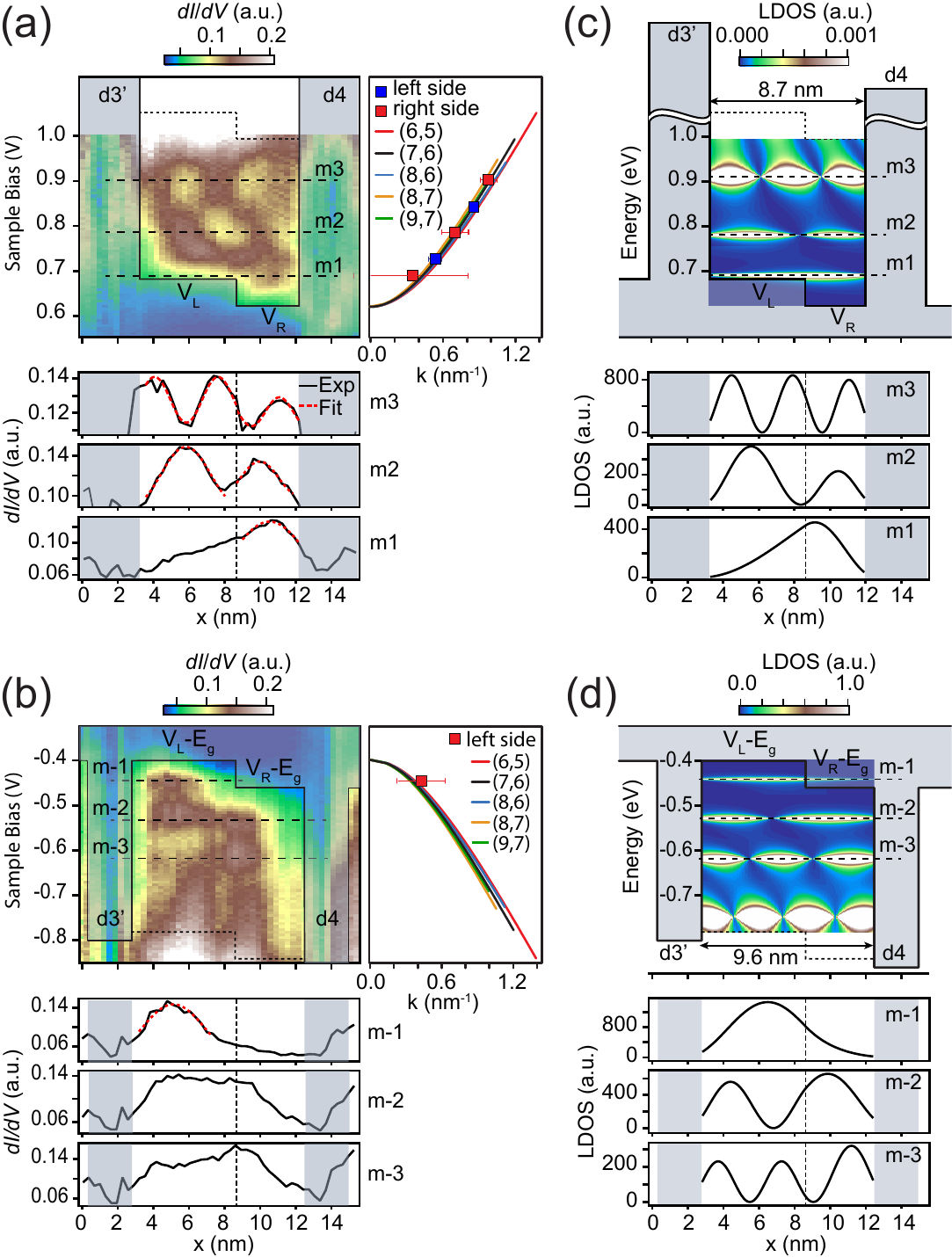}
  \caption{\label{exp_data_Ar} (a)-(b) QD I detailed $dI/dV(x,V)$ maps in conduction and valence bands. Lower subpanels contain QD states linecut profiles and stationary wave-like fits in left and right QD parts. Right subpanels contain experimental energy dispersion relation data sets $k_\mathrm{n}(E_\mathrm{n})$ and tight-binding calculations. (c)-(d) Resulting LDOS calculated from a one-dimensional piecewise constant potential model featuring potential barriers and a potential step (gray area), with position of the potential step: 5.09 nm from the right barrier's center, potential step heigth: $U_\mathrm{C}=V_\mathrm{L}-V_\mathrm{R}=60$ meV, barrier heights: $V_\mathrm{d3'}=1$ eV, $V_\mathrm{d4}=0.85$ eV, barrier widths: $a_\mathrm{d3'}=a_\mathrm{d4}=3.4$ nm.  Valence band: $V_\mathrm{d3'}=-0.4$ eV, $a_\mathrm{d3'}=a_\mathrm{d4}=2.5$ nm, $V_\mathrm{d4}=-0.4$ eV. $E_\mathrm{g}$ stands for bandgap energy.}
\end{figure}
%%%
%
\begin{figure}
  \includegraphics[width=12cm]{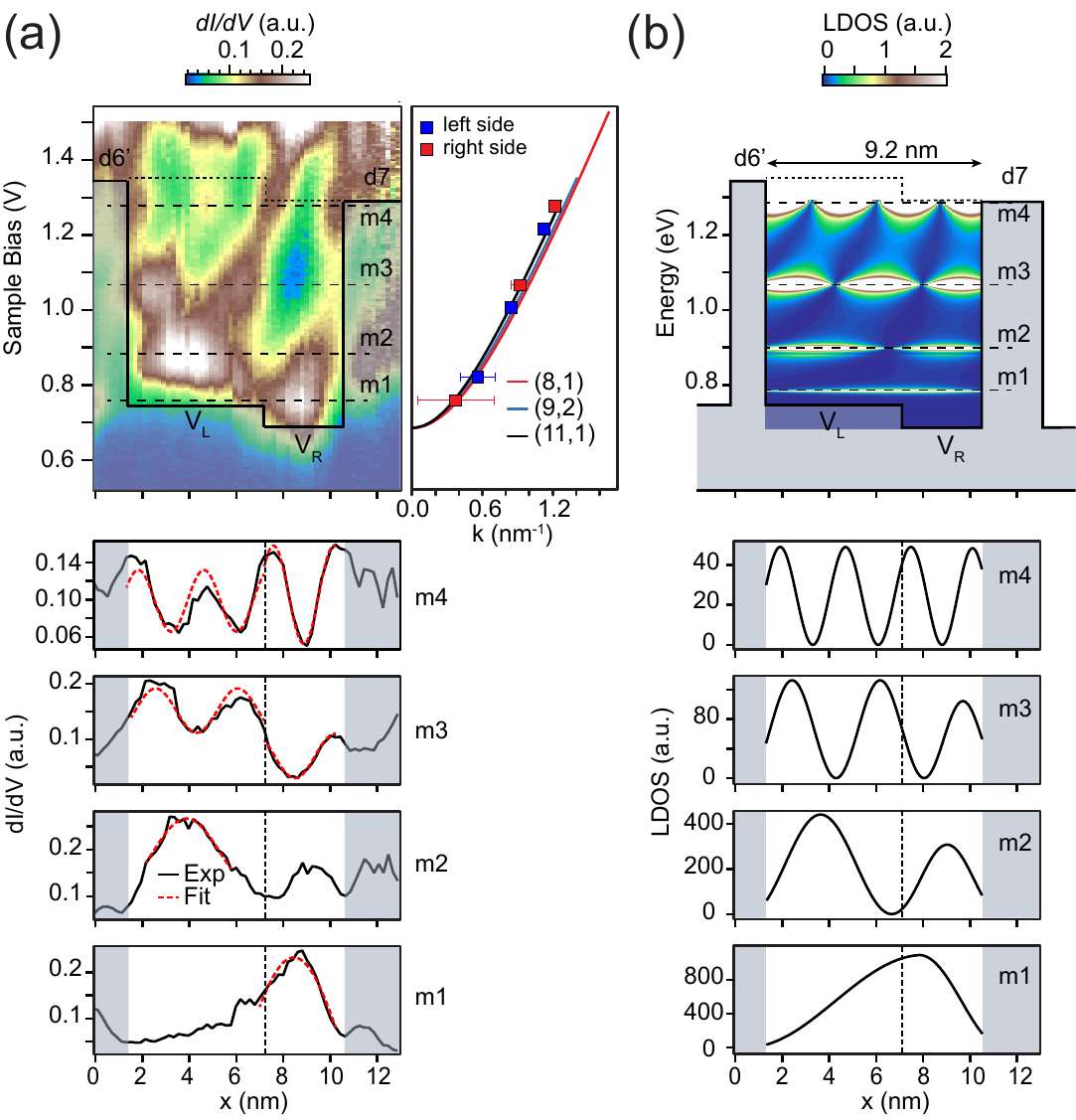}
  \caption{\label{exp_data_N} (a) QD II detailed $dI/dV(x,V)$ map. Lower subpanels contain QD states linecut profiles and stationary wave-like fits in the left and right QD parts. Right subpanel contains experimental energy dispersion relation data sets $k_\mathrm{n}(E_\mathrm{n})$ and tight-binding calculations. (b) Resulting LDOS calculated from a one-dimensional piecewise constant potential model featuring potential barriers and a potential step (gray area) with position of the potential step: 4.7 nm from the right barrier's center, potential step heigth: $U_\mathrm{C}=V_\mathrm{L}-V_\mathrm{R}=60$ meV, barrier heights: $V_\mathrm{d6'}=0.6$ eV, $V_\mathrm{d7}=0.6$ eV, barrier widths: $a_\mathrm{d6'}=1.5$ nm, $a_\mathrm{d7}=2.6$ nm.}
\end{figure}
%%%
%
\indent Remarkably, the $dI/dV(x,V)$ maps in Fig.~\ref{exp_data_1} exhibit several broad discrete states in the conduction bands of SWNT I, II (white dashed boxes in panel (c) and (e), respectively) and in the valence band of SWNT I (white dashed box in panel (c)), characterized by a modulation of the $dI/dV$ signals in the spatial direction between pairs of consecutive defect sites $d3'-d4$ and $d6'-d7$. Enlarged plots of these boxed regions are displayed in Fig.~\ref{exp_data_Ar}(a)-(b) and Fig.~\ref{exp_data_N}(a) for SWNTs I and II, respectively. In the conduction bands, cross-sectional curves recorded along the black horizontal dashed lines labelled m1--m3 in Fig.~\ref{exp_data_Ar}(a) and m1--m4 in Fig.~\ref{exp_data_N}(a) are plotted below the LDOS panels. These clearly reveal one to three and respectively one to four spatially equidistant maxima. The number of maxima increases for increasing $\left|V_\mathrm{bias}\right|$ and the measured level spacings between consecutive discrete states is of the order of 100 meV and larger for both cases. This indicates that defect sites $d3'-d4$ and $d6'-d7$, respectively separated by 12.1 nm and 11.2 nm, act as strong scattering centers able to confine carriers in semiconducting SWNTs~\cite{Buchs_PRL,Bercioux_prb_2011}. Such intrananotube QD structures will be referred as QD I (in SWNT I) and QD II (in SWNT II) in the following. We estimated the level spacings in the conduction band of QD I to 98 meV (m1-m2) and 116 meV (m2-m3). For QD II, we measured 122 meV (m1-m2), 185 meV (m2-m3) and 210 meV (m3-m4).
\\
\indent In the valence band of SWNT I, discrete states with level spacings of the order of 80-90 meV, with one clear maximum at the level m-1, can also be distinguished between defect sites $d3'-d4$ in Fig.~\ref{exp_data_Ar}(b). The discretization of the states indicates that this QD structure also confines holes. Discrete states starting from m-2 and lower show less well defined structures compared to the conduction band states. In the case of SWNT II, no clear discrete states are observed in the valence band (see supplementary information). These observations are most probably the result of an energy dependent scattering strength of the defects, respectively $d3'$-$d4$ and $d6'$-$d7$, leading here to a weaker confinement in the valence band. Such energy dependence is well known for metallic SWNTs~\cite{Chico96,vac_2007,mayrhofer:2011,Bockrath_Science01} and is corroborated by our ab-initio calculations. Note that mixing effects with defect states and substrate-induced effects~\cite{substrate_effects} cannot be ruled out.
\\
\indent Another remarkable feature in the LDOS is the strong spatial asymmetry of the lowest energy states m1 and m-1 in QD I and m1 in QD II. In QD I, m1 is shifted to the right side of the dot while m-1 is shifted to the left side. Higher states m2 and m3 show more symmetry in terms of position of the maxima relative to the center of the QD. In QD II, m1 is shifted to the right side of the QD. We attribute the observed lowest energy states asymmetry (for electrons as well as for holes) in part to their strong sensitivity to weak potential modulations within the QD structure (as we will show in section \ref{1D}). For QD I, this assertion is supported by the observation of a 0.25 nm high Au(111) terrace edge located around the center of the QD, leading to a supported-suspended interface (see white dashed lines in Fig.~\ref{exp_data_1}(b) and more topographic details in Fig.~S2(a)-(d) in supplementary information). Such configurations have been reported to induce a rigid shift in the SWNT bands~\cite{Clair_2011}, for instance here a down-shift in the right side of QD I corresponding to the "suspended" portion between two terraces. In QD II, we attribute the spatial shift of m1 to a potential modulation induced by a layer of disordered impurities, most probably residua from the 1,2-dichloroethane suspension, lying between the gold substrate and the SWNT (see Fig.~\ref{exp_data_1}(d) and Fig.~S2(e)-(h) in supplementary information). 
\\
\indent Also, the LDOS in QD I and II (Fig.~\ref{exp_data_Ar}(a) and Fig.~\ref{exp_data_N}(a), respectively) reveals asymmetric patterns with curved stripes oriented from top left to bottom right for QD I and from bottom left to top right for QD II. These are characteristic signatures for defect pairs with different scattering strengths~\cite{Bercioux_prb_2011,Buchs_PRL}. For instance here, the left defect in QD I ($d3'$) has a larger scattering strength than the right one ($d4$), while the right defect in QD II ($d7$) has a larger scattering strength than the left one ($d6'$). 
\\
\indent The exact atomic structure of the defects could in principle be determined from a comparison of $dI/dV$ spectra with simulated first-principle LDOS signatures of expected defect types. In reality, this is hampered by the large number of possible geometries to simulate, including complex multiple defect structures~\cite{Buchs_Ar}, together with the large unit cells of the semiconducting chiral SWNTs studied here.
\\
\subsection{1D piecewise constant potential model}
\label{1D}
To better understand the physical origins of the non-trivial signatures of the quantized states, we model the experimental $dI/dV$ maps by solving the time independent one-dimensional Schr\"odinger equation over a piecewise constant potential model of QD I and QD II. The scattering centers are approximated by semi-transparent rectangular tunneling barriers leading to a square confinement potential~\cite{Laird:2015}. This is supported by previous results on defect-induced confinement in metallic SWNTs using the same experimental conditions~\cite{Buchs_PRL} and is consistent with ab-initio simulations presented later in this work. The potential modulation within the QD is approximated by a potential step. The resulting potential geometries are illustrated with gray shaded areas in Fig.~\ref{exp_data_Ar} (c) and (d) and Fig.~\ref{exp_data_N}(b). Dispersion relations $E(k)$ can be extracted experimentally from the quantized states wavefunctions by measuring the energy and corresponding momenta in the left and right sides of the QDs. The wavevectors $k$ are determined using stationary wave-like fitting functions~\cite{Buchs_PRL} displayed with dashed red curves in Figs.~\ref{exp_data_Ar}(a)-(b) and ~\ref{exp_data_N}(a)). From this procedure, the potential step height and position can be estimated (see supplementary information). The experimental data sets $E(k)$ are plotted in the right panels of Figs.~\ref{exp_data_Ar}(a) and \ref{exp_data_N}(a) together with dispersion relations from a third-nearest neighbor tight-binding calculation closely approximating ab-initio results~\cite{Reich_TB_2002}. These chirality-dependent tight-binding dispersion relations, calculated within an extended Brillouin zone resulting from the defect-induced breaking of the translation invariance~\cite{Bercioux_prb_2011}, are used in the Hamiltonian of our one-dimensional model. Taking into account the measured chiral angle, diameter distribution~\cite{Buchs_conf} and measured bandgaps, we find the best match with chiralities $(7,6)$ for QD I and $(11,1)$ for QD II (see supplementary information). 
\\
\indent Once chiralities together with potential step heights and positions are optimized, one can fit the height and width of the rectangular tunneling barriers in order to reproduce the experimental level spacings and general LDOS patterns. On a qualitative ground, a symmetric double barrier system results in the formation of spatially symmetric discrete bound states. Increasing both barrier heights simultaneously shifts the bound state energy levels and level spacings up. This leads to sharper bound states as the confinement in the QD is made stronger thus increasing the lifetime of the confined electrons. Increasing the barrier thickness with constant inner edge separation does not affect much the level spacings but further sharpens the bound states. Any asymmetry introduced by a change in the width or height of one single barrier leads to broader bound states. The presence of a potential step modifies the LDOS in lifting the levels of the bound states, with a more pronounced effect on the lower states. In QD I and II, the center of each barrier is aligned with the center of the gap states ($d3'$-$d4$ for QD I and $d6'$-$d7$ in QD II) and the width ratio is kept proportional to the ratio of the spatial extent of the gap states. Thus, by increasing the width of the barriers, we decrease the length of the QD leading to higher level spacings, and vice versa. The experimental level spacings can then be approximated by tuning both barrier widths in the same ratio and the heights individually, knowing that the scattering strength of $d3'$ ($d7$) is larger than $d4$ ($d6'$) according to the observed asymmetry in the LDOS described above \footnote{The transmission probability through a rectangular tunneling barrier is given by $T=\left( 1+\frac{V^{2}\sinh^{2}\left( a \cdot \sqrt{2m^{*}(V-E)}/\hbar \right)}{4E(V-E)} \right)^{-1}$, where $V$ and $a$ are respectively the barrier height and width. For the argument in the $\sinh$ sufficiently small such that $\sinh(x)\simeq x$, it can be shown that $a$ and $V$ can be coupled such that the transmission probability becomes a function of the area under the barrier $A=a\cdot V$, with $T=\left( 1+ \frac{m^{*}A^{2}}{2\hbar^{2}E} \right)^{-1}$. In our case, this condition is not satisfied and thus the barrier geometries are tuned empirically to fit the experimental level spacings.}. 
\\
\indent For QD I, we find a good match in the conduction band for the barrier heights $V_\mathrm{d3'}=1$ eV and $V_\mathrm{d4}=0.85$ eV, widths $a_\mathrm{d3'}=a_\mathrm{d4}=$ 3.4 nm,  and potential step $V_\mathrm{L}-V_\mathrm{R}=60$ meV. With these parameters, the spatial profile of the obtained quantized states (see lower subpanels in Fig.~\ref{exp_data_Ar}(a) and (c)) reproduces the experimental modulation features remarkably well. Also, the simulated LDOS displays a pattern with curved stripes oriented from top left to bottom right, as observed experimentally, due to a left barrier with a larger scattering strength. In the valence band, although modes m-2 and lower do not show a well defined structure in the spatial direction, thinner barriers with dimensions $a_\mathrm{d3'/d4}=2.5$ nm, $V_\mathrm{d3'/d4}=-0.4$ eV, leading to a slightly longer QD length (9.6 nm compared to 8.7 nm in the conduction band) can reproduce the measured level spacings very well. 
\\
\indent For QD II, we observed that the measured energy levels are overestimated by a factor $\alpha\sim1.29$, presumably due to a voltage division effect induced by the impurity layer mentioned above (see details in supplementary information). We find a good agreement with the experimental LDOS with the parameters: $V_{d3'}=V_{d4}\simeq$ 0.47 eV, $a_\mathrm{d6'}=1.5$ nm, $a_\mathrm{d7}=2.6$ nm and $U_\mathrm{C}=V_\mathrm{L}-V_\mathrm{R}\simeq 47$ meV. Note that in Fig.~\ref{exp_data_N}(b) the barrier and potential heights are multiplied by $\alpha$ to allow a direct comparison with the experimental LDOS. The simulated LDOS shows a pattern with curved stripes oriented from bottom left to top right, as observed experimentally, due to a right barrier exhibiting a larger scattering strength. Also, the spatial profile of the obtained bound states (see lower subpanels in Fig.~\ref{exp_data_N}(a) and (b)) reproduces the experimental features quite well. Note also that one can distinguish an isolated state in the experimental LDOS at an energy level between m1 and m2, about in the middle of the QD. This state that prevented an accurate fit of the state m2 in the right QD part is attributed to a spatial feature visible in the STM topography image in Fig.~\ref{exp_data_Ar}(d) (see also supplementary information, Fig.S2(f)), probably a physisorbed impurity which does not affect the LDOS significantly.
\\
\subsection{Ab-initio calculations}
%
%%%
\begin{figure}
  \includegraphics[width=16cm]{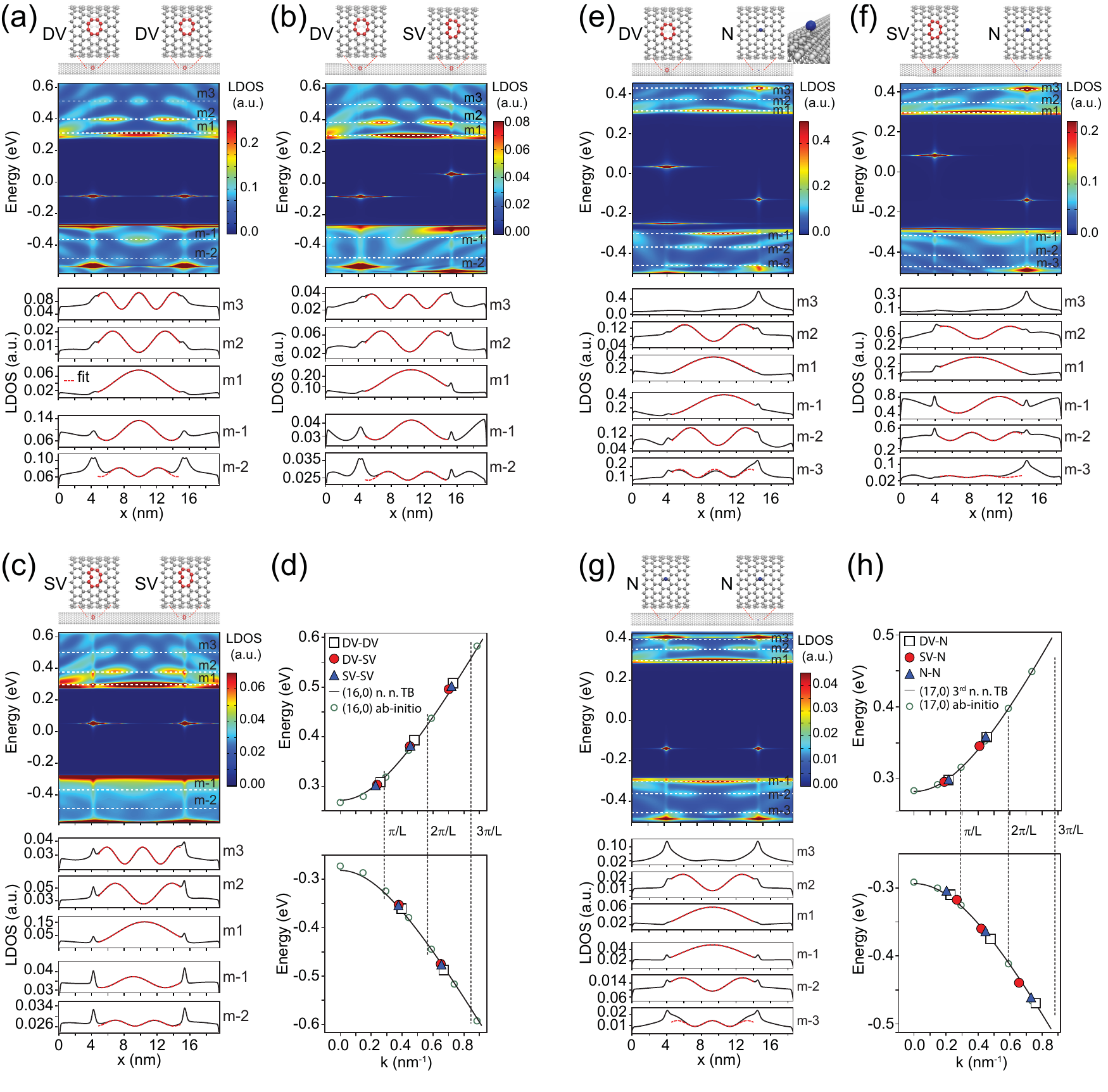}
  \caption{\label{num_data} (a)-(c) LDOS ab-initio simulations of a semiconducting $(16,0)$ SWNT with combinations of vacancies defects separated by 11.1 nm. Subpanels display QD state linecut profiles. (d) Tight-binding (black curve) and ab-initio dispersion relations (green circles) for a pristine $(16,0)$ SWNT with $E_\mathrm{n}(k_\mathrm{n})$ data sets extracted from (a)-(c). (e)-(g) LDOS ab-initio simulations of a semiconducting $(17,0)$ SWNT with combinations of N ad-atoms and vacancies defects separated by 10.7 nm. (h) Tight-binding (black curve) and ab-initio dispersion relations (green circles) for a pristine $(17,0)$ SWNT with $E_\mathrm{n}(k_\mathrm{n})$ data sets extracted from (e)-(g).}
\end{figure}
%%%
%
In order to elucidate the physical nature of the electron/hole confining scattering centers, we performed ab-initio simulations based on a combination of density functional theory~\cite{pbe,paw,vasp_paw,VASP2}, maximally localized Wannier orbitals~\cite{transportwannier90} and Green's functions (see supplementary information). Without loss of generality, we have simulated short unit cell semiconducting zigzag SWNTs with different combinations of the most probable defect structures. Results for vacancy defects likely being induced by 200 eV Ar$^{+}$ ions, separated by about 11 nm in a $(16,0)$ SWNT are shown in Fig.~\ref{num_data}(a)-(c) with DV-DV, DV-SV and SV-SV pairs, respectively. The LDOS displays midgap states at the defect positions as expected as well as defect states in the valence band~\cite{Buchs_Ar}. Most importantly, clear quantized states with a number of maxima increasing with energy are observed between the defects in the conduction band, emphasizing the ability of SVs and DVs to confine carriers. For the asymmetric configuration DV-SV, one can distinguish faint curved stripe patterns oriented from top left to bottom right, indicating a larger scattering strength for DVs compared to SVs. This is consistent with observations in transport experiments~\cite{Gomez05nm}. On the other hand, the patterns in the valence band strongly depend on the defect types. Discrete states can be distinguished for the DV-DV case, with m-2 being mixed with defect states. For the DV-SV case, clear curved stripe patterns oriented from bottom left to top right indicate again a stronger scattering strength for DV. Also, broader states are observed, indicating that the scattering strength of DVs and SVs is weaker in the valence band compared to the conduction band.
\\
\indent More insight on the energy dependent scattering strength for each defect pair configuration can be obtained by extracting the wavevector $k_\mathrm{n}(E_\mathrm{n})$ for each resonant state. This data set is plotted in Fig.~\ref{num_data}(d) for the conduction and valence bands together with the $(16,0)$ dispersion relations calculated from the third-nearest neighbor TB model and from the ab-initio calculation for the pristine nanotube. A first observation is the excellent agreement between TB and ab-initio results, further validating the method used in Figs.~\ref{exp_data_Ar}(a)-(b) and ~\ref{exp_data_N}(a). The vertical dashed lines indicate the limiting $k_\mathrm{n,\infty}=\frac{\pi \cdot n}{L}$ values corresponding to the closed system (infinite hard walls potential) with $L=11.1$ nm being the defect-defect distance. In the conduction band, we find that $k_\mathrm{n}(E_\mathrm{n})=\frac{\pi \cdot n}{L_\mathrm{eff}(n)} < k_\mathrm{n,\infty}$, indicating that the effective lengths $L_\mathrm{eff}(n)$ of the QD are larger than $L$ ($i.e.$ the resonant states wavefunctions are characterized by penetrating evanescent modes inside the defect scattering potential), as expected for an open system. The shortest $L_\mathrm{eff}(n)$ are obtained for the DV-DV configuration with 12.1 nm (m1), 13.1 nm (m2) and 12.9 nm (m3), which we attribute to wider scattering potential profiles for DVs compared to SVs. In the valence band, we find that $k_\mathrm{n}(E_\mathrm{n})=\frac{\pi \cdot n}{L_\mathrm{eff}(n)} > k_\mathrm{n,\infty}$, with $L_\mathrm{eff}(n)$ values between 7.9 nm (DV-DV, m-1) and 9.66 nm (DV-SV, m-2). We attribute this pronounced QD shortening to wider scattering potential profiles of both DVs and SVs in the valence band, probably due to mixing with wide spread defect states in the valence band.
\\
\indent Ab-initio calculations for different defect pairs combinations containing at least one N ad-atom, $i.e.$ N-DV, N-SV and N-N, are presented in Fig.~\ref{num_data}(e)-(h) for a $(17,0)$ SWNT, along with details on the defects geometries. Remarkably, clear QD states are generated for all three configurations, underlining the potential of N ad-atoms to confine carriers in semiconducting SWNTs and thus to generate intrananotube QDs. 
\\
\indent In order to demonstrate the scattering strengths of the different defects, we calculated the energy dependent conductance in addition to the LDOS for the different combinations of the QD defining scattering defects on the $(16,0)$ and $(17,0)$ SWNTs, see supplementary information. Generally we can observe strong conductance modulation of the order of 30-40\% with regard to the pristine CNT for all three tested defects (double vacancies DV, single vacancies SV and chemisorbed C-N) with the DVs having the largest scattering strength in the CB and VB.  
\\
\indent Note that the choice of the zigzag SWNT chiralities in the two different ab-initio scenarios is motivated by the different effective masses of both chiralities ($m^{*}_{(17,0)}>m^{*}_{(16,0)}$) which is typical for chirality families $(3n-1,0)$ and $(3n-2,0)$~\cite{ZZ_families}. Taking advantage of recent reports on SWNT chirality control~\cite{chirality_control_EMPA,chirality_control_chinese,chirality_chemistry}, this property could be used in practice to design QDs with different level spacings for the same QD length. From an application point of view, however, QDs generated by DVs will have far superior stability at room temperature due to their high migration barrier above 5 eV ($\sim$~1 eV for single vacancy)~\cite{Kra06vm}. This value drops down by at least 2 eV for N ad-atoms depending on their chemisorption configuration~\cite{Nitrogen_prb_07,Yma05nitr}.
\\
\indent Our ab-initio simulations do not take into account any substrate effect. In the experimental case, the carriers can decay through the substrate, thus limiting their lifetime. This leads to state broadening, measured between about 60 meV up to 120 meV in QD I and II, while the quantized states widths in ab-initio simulations vary between about 5 meV and 45 meV. This suggests that a better contrast of the experimental quantized states, especially in the valence band, could be achieved by lowering the nanotubes-substrate interaction through $e.g.$ the insertion of atomically thin insulating NaCl films~\cite{Ruffieux_Nature_2016}. This would allow to gain more insight on the electronic structure of the QDs as well as in the associated scattering physics at the confining defects~\cite{Buchs_PRL}. 

\section{Conclusions and outlook}
In summary, using low-temperature STM/STS measurements supported by an analytical model and ab-initio simulations, we have demonstrated that intrananotube quantum dots with confined electron and hole states characterized by energy level spacings well above thermal broadening at room temperature can be generated in semiconducting SWNTs by structural defects such as vacancies and di-vacancies, as well as nitrogen ad-atoms. These results, combined with recent progresses in type and spatial control in the formation of defects~\cite{Robertson_2012,Yoon_2016,Laser_writing_2017} as well as chirality control~\cite{tunable_QD_defects}, hold a high potential for applications in the design of SWNT based quantum devices. These include $e.g.$ electrically driven single-photon emitters operating at room temperature and telecom wavelength. In this context, the observation of quantum confinement effects in the emitted light of cut, sub-10 nm, semiconducting SWNTs~\cite{Dai_2008} shall be seen as an additional motivation for investigating the optical properties of our "QD with leads" building-blocks. These would include $e.g.$ studying optical transitions selection rules for different types and configurations of defect pairs~\cite{sel_rules_2006} associated with experimental studies such as photoluminescence~\cite{Lefebvre06} combined to $g^{(2)}$ correlation measurements~\cite{Hofmann_2013} in suspended SWNT devices as well as photocurrent imaging~\cite{Buchs_Nat_comm} and spectroscopy~\cite{Gabor_2009}.

\section*{Acknowledgements}
The authors thank Ethan Minot, Lee Aspitarte, Jhon Gonzalez, Andres Ayuela, Omjoti Dutta and Arkady Krasheninnikov for fruitful discussions.
The work of DB is supported by Spanish Ministerio de Econom\'ia y Competitividad (MINECO) through the project  FIS2014-55987-P and by the (LTC) QuantumChemPhys. LM acknowledges support from the BMBF-project WireControl (FKZ16ES0294) and computing time for the supercomputers JUROPA and JURECA at the J\"ulich Supercomputer Centre (JSC).

\clearpage

\section*{References}

%\bibliography{cnt}

\end{document}